\documentclass[superscriptaddress,groupedaddress,12pt]{article}  
\pdfoutput=1
\usepackage{graphicx}  
\usepackage{dcolumn}   
\usepackage{bm}        
\usepackage{amssymb}   
\usepackage{jcappub}

\usepackage{bm}
\usepackage{latexsym}
\usepackage{amsmath}
\usepackage{amsfonts}
\usepackage{colortbl}
\usepackage{multirow}
\usepackage{array}
\usepackage{booktabs}
\usepackage{rotating}
\usepackage{epstopdf}
\usepackage{subfigure}

\renewcommand{\arraystretch}{1.5}

\begin{document}

\title{Shining Light on Modifications of Gravity}

\author[1]{Philippe Brax,}
\author[2]{Clare Burrage}
\author[3]{and Anne-Christine Davis}
\affiliation[1]{Institut de Physique Th\'{e}orique, CEA, IPhT, CNRS, URA2306, F-91191
Gif-sur-Yvette c\'{e}dex, France}
\affiliation[2]{School of Physics and Astronomy, University of Nottingham, Nottingham NG7 2RD, UK}
\affiliation[3]{Department of Applied Mathematics and Theoretical Physics,
Centre for Mathematical Sciences, Cambridge CB3 0WA, UK}

\emailAdd{Philippe.Brax@cea.fr}
\emailAdd{Clare.Burrage@nottingham.ac.uk}
\emailAdd{A.C.Davis@damtp.cam.ac.uk}

\abstract{Many modifications of gravity introduce new scalar degrees of freedom, and in such theories matter fields typically couple to an effective metric that depends on both the true metric of spacetime and on the scalar field and its derivatives.  Scalar field contributions to the effective metric can be classified as conformal and disformal.  Disformal terms introduce gradient couplings between scalar fields and the energy momentum tensor of other matter fields, and cannot be constrained by fifth force experiments because the effects of these terms are trivial around static non-relativistic sources.   The use of high-precision, low-energy photon experiments to search for conformally coupled scalar fields, called axion-like particles, is well known.  In this article we show that these experiments  are also constraining for disformal scalar field theories, and are particularly important  because of the difficulty of constraining these couplings with other laboratory experiments.
 }

\maketitle

\section{Introduction}
Light scalar fields are both extremely common and extremely annoying in the study of cosmology and modifications of gravity.  They are often invoked as quintessence fields to explain the late time acceleration of the expansion of the universe, and in many modifications of gravity, from $f(R)$ theories, through DGP to theories of massive gravity additional light scalar degrees of freedom appear in the gravity sector.  The existence of new dynamical degrees of freedom  makes these theories distinguishable from standard general relativity and $\Lambda$CDM cosmology.  However the  consequence of introducing new light scalar degrees of freedom is that the scalar field will mediate a new long range fifth force and  we know this to be incompatible with experimental tests to a high degree of precision.

In such theories Standard Model matter fields feel an effective metric  $\tilde{g}_{\mu\nu}$ that depends on the true metric $g_{\mu\nu}$, which determines the geometry of  space, and on the scalar fields and their derivatives.  It was shown by Bekenstein \cite{Bekenstein:1992pj} that the most general metric that can be constructed from  $g_{\mu\nu}$ and a scalar field, and  that respects causality and the weak equivalence principle is\footnote{In his analysis Bekenstein excludes the possibility of including scalars with more than one derivative apiece because they were expected to give rise to ghost degrees of freedom.  Recent developments, including the Galileon models \cite{Nicolis:2008in}, have shown that higher derivative terms are not automatically problematic, so it is possible that Bekenstein's metric can be extended in this way.  However no such extensions are currently known so we concentrate our analysis on the metric in Equation (\ref{eq:bekenstein}).}
\begin{equation}
\tilde{g}_{\mu\nu}=A(\phi,X)g_{\mu\nu} +B(\phi,X)\partial_{\mu} \phi\partial_{\nu}\phi\;,
\label{eq:bekenstein}
\end{equation}
where $X=-(\partial \phi)^2/2$, and $|B|=-B$.  The metric (\ref{eq:bekenstein}) now defines the geodesics and light cones for the particles of the Standard Model.  The first term in this expression is known as the conformal term, and the case $A\equiv A(\phi)$ and $B\equiv 0$ has been much studied. For light, $m<(1\mbox{ mm})^{-1}$, conformally coupled scalars laboratory searches for a fifth force constrain the strength of the coupling to be $M\gtrsim10^5 M_P$ \cite{Adelberger:2009zz}.  This transplanckian energy scale is problematic if we want to construct a natural quantum theory for such scalar fields and is a new fine tuning issue for any such theory.  Conformally coupled  scalar fields can be experimentally acceptable without fine tuning however if we allow for the theory to be non-linear.  Non-linear mass terms (in the Chameleon model \cite{Khoury:2003rn,Brax:2004qh}), non-linear forms of the coupling function $A(\phi)$ (in the symmetron  \cite{Pietroni:2005pv,Olive:2007aj,Hinterbichler:2010es} and dilaton \cite{Damour:1994zq,Brax:2010gi} models),  and non-linear kinetic terms (in the Galileon model \cite{Nicolis:2008in,Deffayet:2009wt} which has a Vainshtein mechanism \cite{Vainshtein:1972sx,Deffayet:2001uk}) can all have the consequence that although the scalar mediated fifth force is large in vacuum it is dynamically suppressed in dense environments, and so does not give rise to noticeable effects in experimental tests.

The second term in equation (\ref{eq:bekenstein}) is the disformal term.  On its own, when $A\equiv\mbox{ const}$, this term is not problematic for fifth force
 experiments because as we will see a scalar field that couples to matter in this way is not sourced by a static non-relativistic mass distribution.  All of the objects used in searches for fifth forces are indeed static and non-relativistic, and so such experiments cannot probe these couplings.

  Scalar fields that give rise to disformal metrics have appeared in a number of circumstances including:
\begin{itemize}
\item In theories of massive gravity \cite{deRham:2010ik}.  A massive graviton can be decomposed into one helicity-two mode, two helicity-one modes and a helicity-zero mode.  Matter couples to a metric
\begin{equation}
h_{\mu\nu}=\hat{h}_{\mu\nu}+\pi\eta_{\mu\nu}+\frac{(6c_3-1)}{\Lambda^3_3}\partial_{\mu}\pi\partial_{\nu}\pi\; ,
\end{equation}
where $\hat{h}_{\mu\nu}$ and $\pi$ are the canonically normalised helicity-two and helicity-zero modes respectively\footnote{We have quoted the standard expression for massive gravity where the fields are expressed in a dimensionless way (this is also true of the DBI-Galileon case that follows).  This is in contrast to the expressions we give in the rest of this article where the scalar fields have dimensions of mass.}.  $c_3$ is an order one dimensionless constant and $\Lambda^3_3= M_Pm^2$, where $m$ is the mass of the graviton\footnote{To be precise this is the metric  in the low energy decoupling limit of the theory where $m\rightarrow 0$, $M_P\rightarrow \infty$ keeping $\Lambda_3=(M_Pm^2)^{1/3}$ fixed.}.

\item In the probe brane world construction of the unified DBI-Galileon model \cite{deRham:2010eu}, where the scalar field $\pi$ describes the location of the brane in a flat bulk space time.  The metric that is induced on the brane is
\begin{equation}
g_{\mu\nu}=e^{-2\pi/l}\eta_{\mu\nu}+\partial_{\mu}\pi\partial_{\nu}\pi \;.
\end{equation}
\item Disformal couplings arise in theories in which  Lorentz invariance is broken spontaneously on a non-trivial background \cite{Brax:2012hm}.
\item The effects of including disformal couplings in chameleon theories were considered in \cite{Noller:2012sv}.
\item Scalar fields with disformal couplings have  been used to give rise to unusual forms of cosmic acceleration, called disformal inflation \cite{Kaloper:2003yf} in the early universe and disformal quintessence \cite{Koivisto:2008ak} in the late universe.  The absence of fifth forces for a disformal  quintessence model was discussed in \cite{Koivisto:2012za}.
\item Disformal scalar fields have also been used previously to give rise to varying speed of light cosmologies \cite{Clayton:1999zs,Drummond:1999ut}, although these are now in some tension with observations \cite{Magueijo:2003gj}.
\end{itemize}
These theories have many other interesting properties, but in this article we are only interested in the consequences of a disformal coupling to matter.  As we have said previously, fifth force experiments do not constrain disformal couplings, however in models motivated by Galileon theories and massive gravity some constraints have been put on the theories from studying gravitational lensing and the velocity dispersion of galaxies \cite{Wyman:2011mp,Sjors:2011iv}.

Fluctuations of these disformally coupled scalar fields do not propagate on geodesics of the metric (\ref{eq:bekenstein}).  Instead their motion is described by a metric that depends on the geometric metric $g_{\mu\nu}$, on the background configuration of the scalar field and on the background distribution of matter fields.  Understanding how disformally coupled scalar fields propagate on non-trivial backgrounds is one of the aims of this paper.  If the scalar field also has non-canonical kinetic terms, as in the massive gravity and Galileon cases, then these terms give additional contributions to the metric describing the propagation of scalar fluctuations, detailed discussion of this point can be found in references \cite{Babichev:2007dw,Burrage:2011cr}.

Fifth force searches are not the only way to study  new scalar fields.  Another extremely fruitful approach is to look for the effects of the mixing of these scalar fields with photons.  Until now the focus of this study has been on  scalar fields that couple without derivative terms, which are referred to as axion-like particles or ALPs because their coupling to photons has the same form as that of the Pecci-Quinn axion.  Conformally coupled scalars are  perfect examples of axion-like particles\footnote{The kinetic term for photons is actually conformally invariant, and so at first glance it seems that a conformally coupled scalar field should not be an axion-like particle.  However as the Standard Model is not conformally invariant a coupling between the scalar field and photons is always generated \cite{Brax:2010uq}.}.    In the presence of a magnetic field photons can convert into these scalar fields and vice versa \cite{Raffelt:1987im}.  If the probability of this conversion is small then this generates rotation and ellipticity of the polarisation of the light beam, if the probability of conversion is large it can lead to the dimming of a light beam.  The mixing can also lead to more exotic effects such as  `light shining through walls', \cite{Redondo:2010dp}, where a magnetic field is used to convert photons into scalar fields which can pass through solid objects that would be impermeable to photons, a second magnetic field on the far side of the wall is used to convert the scalar field back into a photon giving the appearance that the light has travelled through the wall.  Currently the strongest experimental constraints on the effects of scalars on the polarisation of photons are from the PVLAS experiment at the Laboratori Nazionali di Legnaro \cite{Zavattini:2007ee}, and  for light shining through walls the best constraints come from the ALPS experiment at the Deutsches Elektronen Synchrotron DESY \cite{Ehret:2010mh}. A nice review of this subject is given in  \cite{Jaeckel:2010ni}.

For canonically coupled scalar fields the constraints of photon laboratory experiments  either looking for changes in the polarisation of light shone through a magnetic field or for light shining through wall effects are far from being as stringent as those from fifth force experiments.  However their potential for constraining disformally coupled scalar fields has so far not been considered, this is the aim of this paper.

In the next section we begin by setting up the disformal scalar field theory we intend to study, and reviewing in Section \ref{sec:fifthforces} why there are no constraints on disformal couplings from fifth force experiments.  In Section \ref{sec:photons} we show how such a scalar field couples to photons and derive the coupled equations of motion for photons and scalars.  In Section \ref{sec:magneticfield} we specialise to the case where photons and scalar fields are propagating through a homogeneous constant background magnetic field, and in Section \ref{sec:eom} we solve the equations of motion.  In Section \ref{sec:labconstraints} we derive the constraints that photon experiments place on disformally coupled scalar fields, for the ALPS experiment in Section \ref{sec:ALPS} and for the PVLAS experiment in Section \ref{sec:PVLAS}.  We conclude in Section \ref{sec:conc}.

We take the signature of the metric to be $(-+++)$.

\section{Disformally coupled scalar fields}

To investigate the effects of a disformally coupled scalar field on the propagation of photons we consider the following effective metric.
\begin{equation}
\tilde{g}_{\mu\nu}=\left(1+\frac{\phi}{\Lambda}\right)g_{\mu\nu}+\frac{2}{M^4}\partial_{\mu}\phi\partial_{\nu}\phi \;.
\end{equation}
Where we have also allowed for a conformal term so that we can most easily highlight the differences in phenomenology between conformally and disformally coupled scalar fields.  $M$ and $\Lambda$ are  new energy scales that, for the moment, we will leave arbitrary. In some cases $\Lambda$ and $M$ will be of the same order of magnitude, but for completeness we will treat them separately.  This is clearly not as general as the metric in equation (\ref{eq:bekenstein}), but our interest is in the disformal term as it has not been studied in this context before,
 and constraints on this term cannot be inferred from previous results. 
 We will assume that the scalar field terms give small corrections to the standard picture, and therefore we have neglected terms that are suppressed by $1/(\Lambda M^4)$.

\subsection{No fifth forces constraints for disformally coupled scalars}
\label{sec:fifthforces}
We begin by reviewing  why there are no fifth force constraints on disformally coupled scalars.  If there is only the disformal term and no conformal term then the situation is particularly simple.  The effective metric is
\begin{equation}
\tilde{g}_{\mu\nu}=g_{\mu\nu}+\frac{2\partial_{\mu}\phi\partial_{\nu}\phi}{M^4}\; ,
\end{equation}
 and the Lagrangian   for a massive scalar field coupled in this way is
\begin{equation}
\mathcal{L}_{\phi}=-\frac{1}{2}(\partial\phi)^2-\frac{1}{2}m^2\phi^2+\frac{1}{M^4}\partial_{\mu}\phi\partial_{\nu}\phi T^{\mu\nu}\; ,
\end{equation}
 resulting in the  scalar equation of motion
\begin{equation}
\Box \phi-m^2 \phi -\frac{2}{M^4}\nabla_{\mu}(\partial_{\nu}\phi T^{\mu\nu})=0\;.
\end{equation}
For fifth force experiments the massive sources are non-relativistic, static and spherically symmetric giving $T^{\mu\nu}=\mbox{diag}(\rho(r),0,0,0)$ and any resulting scalar field profile will also be static and spherically symmetric $\phi\equiv \phi(r)$.  In such a  scenario it is clear that the source term vanishes and the scalar field equation of motion is trivial. If a massive object does not source a scalar field profile then there is no resulting fifth force.

If there is also a conformal term present the metric is
\begin{equation}
\tilde{g}_{\mu\nu}=\left(1+\frac{\phi}{\Lambda}\right)g_{\mu\nu}+\frac{2}{M^4}\partial_{\mu}\phi\partial_{\nu}\phi\;,
\label{eq:metric}
\end{equation}
and  for a massive scalar field the equation of motion is
\begin{equation}
\Box\phi -m^2\phi +\frac{T}{\Lambda}-\frac{2}{M^4}\nabla_{\mu}(\partial_{\nu}\phi T^{\mu\nu})=0\;.
\label{eq:boxphi}
\end{equation}
As before, for a static, non-relativistic  source the last term in (\ref{eq:boxphi}) vanishes, but now the scalar field is sourced by the $T/\Lambda$ term in the equation of motion.  When $mr\ll1$, and when the source is a spherically symmetric object of mass $M_c$  the solution to the equation of motion is
\begin{equation}
\phi\approx \frac{M_c}{\Lambda r}\;.
\end{equation}
 As the scalar field profile is non-trivial a scalar force will be generated.

The fifth forces felt by matter due to the presence of a conformally or disformally coupled scalar field can be determined from the geodesic equation for matter.  Matter fields move on geodesics of the metric $\tilde{g}_{\mu\nu}$ and so their movement is governed by the equation
\begin{equation}
\tilde{u}^{\nu}\tilde{\nabla}_{\nu}\tilde{u}^{\mu}=0\;,
\label{eq:geodesic}
\end{equation}
where $\tilde{u}^{\mu}$ is the four velocity of particles normalised with respect to the tilded metric $\tilde{g}_{\mu\nu}\tilde{u}^{\mu}\tilde{u}^{\nu}=-1$, and $\tilde{\nabla}$ is the covariant derivative with respect to that metric.  If we now express the tilded metric in terms of the true spacetime metric, and define a new four velocity $u^{\mu}\propto\tilde{u}^{\mu}$ such that $g_{\mu\nu}u^{\mu}u^{\nu}=-1$ and an acceleration $a^{\mu}=u^{\nu}\nabla_{\nu}u^{\mu}$ then the geodesic equation (\ref{eq:geodesic}) can be written as
\begin{equation}
a^{\mu}=F^{\mu}(\phi, \partial\phi, \partial\partial\phi)\;,
\end{equation}
where $F^{\mu}$ is the force per unit mass experienced by matter fields due to the presence of a scalar field.
 The general expression for the scalar force  is rather involved, but when the system is static and spherically symmetric and the source is non relativistic the  scalar force is only non-zero in a radial direction and given by
\begin{equation}
F_{r}=-\frac{\phi^{\prime}}{2\Lambda}\left(1+\frac{\phi^{\prime 2}}{M^4 + \phi^{\prime 2}}\right)\;,
\end{equation}
where $\prime$ denotes differentiation with respect to $r$.  Corrections which depend on the energy scale of the disformal term $M^4$ are always subleading, and therefore not constrained by the null results of fifth-force experiments.

\subsection{The interaction with photons}
\label{sec:photons}

In order to  devise tests for disformally coupled scalar fields we consider how these fields interact with photons.  The Lagrangian describing the scalar fields, photons and their interactions is
\begin{equation}
\mathcal{L}_{\phi,\gamma}=-\frac{1}{2}(\partial\phi)^2 -V(\phi) -\frac{1}{4}F^2 -\frac{\phi}{\Lambda}F^2 -\frac{1}{M^4}\partial_{\mu}\phi\partial_{\nu}\phi \left[\frac{1}{4}g^{\mu\nu}F^2 -F^{\mu}_{\;\; \alpha}F^{\nu\alpha}\right]\;,
\label{eq:lagphotscal}
\end{equation}
where, as usual, $F_{\mu\nu}=\partial_{\mu}A_{\nu}-\partial_{\nu}A_{\mu}$.

The resulting scalar equation of motion is
\begin{equation}
-\Box\phi -\frac{2}{M^4}\nabla_{\mu}\left\{\partial_{\nu}\phi \left[\frac{1}{4}g^{\mu\nu}F^2 -F^{\mu}_{\;\;\alpha}F^{\nu\alpha}\right]\right\}
=-V^{\prime}-\frac{1}{\Lambda}F^2\;,
\end{equation}
and the equation of motion for the photon $A_{\mu}$ is
\begin{equation}
\nabla^{\sigma}\left[F_{\sigma\rho}\left(1+\frac{4\phi}{\Lambda}+\frac{(\partial\phi)^2}{M^4}\right)+\frac{2}{M^4}(F^{\nu}_{\;\;\rho}\partial_{\sigma}\phi\partial_{\nu}\phi-F^{\nu}_{\;\;\sigma}\partial_{\rho}\phi\partial_{\nu}\phi)\right]=0\;.
\end{equation}
In these equations of motions we have neglected higher order interactions that are suppressed by $\mathcal{O}(1/\Lambda M^4)$.  In Lorentz gauge, $\partial_{\alpha}A^{\alpha}=0$, and  in flat space these equations of motion reduce to
\begin{eqnarray}
\Box\phi\left(1+\frac{1}{2M^4}F^2\right)-\frac{2}{M^4}\partial_{\mu}\partial_{\nu}\phi F^{\mu}_{\;\;\alpha}F^{\nu\alpha}& & \\
{}+\frac{2}{M^4}\partial_{\nu}\phi(2\partial^{\nu}\partial_{\alpha}A_{\beta}F^{\alpha\beta}+\Box A_{\alpha}F^{\nu\alpha})&=&V^{\prime}+\frac{1}{\Lambda}F^2 \nonumber\;,
\end{eqnarray}
and
\begin{eqnarray}
0 &=& \Box A_{\rho}+\frac{4}{\Lambda}(\phi\Box A_{\rho}+F_{\sigma\rho}\partial^{\sigma}\phi) \\
& & {}+\frac{1}{M^4}\left[\begin{array}{l}
\Box A_{\rho}(\partial\phi)^2+4F_{\sigma\rho}\partial_{\alpha}\phi\partial^{\sigma}\partial^{\alpha}\phi+2(\partial_{\sigma}F_{\nu\rho})\partial^{\sigma}\phi\partial^{\nu}\phi \\
 {}+2F_{\nu\rho}\Box\phi\partial^{\nu}\phi+2\Box A_{\nu}\partial_{\rho}\phi\partial^{\nu}\phi-2F_{\nu\sigma}(\partial^{\sigma}\partial_{\rho}\phi\partial^{\nu}\phi -\partial_{\rho}\phi\partial^{\sigma}\partial^{\nu}\phi)
 \end{array}\right]\nonumber\;.
\end{eqnarray}

\section{Propagation through a constant magnetic field}
\label{sec:magneticfield}

\subsection{Perturbation equations}

Laboratory searches for the effects of scalars on the propagation of photons consist of a polarised beam of photons  passing through a constant magnetic field oriented perpendicular to the direction of propagation of the photons.  We consider propagating photons and scalars as perturbations about such a background. The background magnetic field is given by
\begin{eqnarray}
A_{0}&=&0\;,\\
A_i&=& \frac{1}{2}\epsilon_{ijk}B_jx_k\;.
\end{eqnarray}
The scalar field background is a constant $\phi_0$ which satisfies
\begin{equation}
V^{\prime}(\phi_0)=-\frac{2}{\Lambda}B^2\;.
\end{equation}
The coupling to photons means that the scalar field is governed by an effective potential
\begin{equation}
V_{\rm eff}(\phi)=V(\phi)+\frac{2\phi B^2}{\Lambda}\;,
\end{equation}
and the background field chooses its value to minimise the effective potential, this is behaviour similar to that of a chameleon scalar field \cite{Khoury:2003rn}.

The theory we are studying is predictive only if  quantum corrections are under control, and do not destroy the Taylor approximation to the form of the conformal coupling.  This requires $\phi_0/\Lambda\ll 1$.
If   the scalar potential is a mass term $V=m^2\phi^2/2$  the background value is
\begin{equation}
\phi_0=-\frac{2B^2}{\Lambda m^2}\;.
\end{equation}
For a light field the background field value may become very large.  As the motivation for studying disformally coupled scalar fields comes from modifications of gravity, we might expect these parameters to be $m\sim H_0$ and $\Lambda\sim M_P$.  A typical experiment will have a Tesla strength magnetic field, this would give
\begin{equation}
\phi_0\sim 10^{14}M_P\;,
\end{equation}
which strongly violates the requirement $\phi_0/\Lambda\ll1 $ for the theory to be safe from quantum corrections.  Obviously it is possible with a different choice of potential or a larger mass of the field to ensure that the background value of the scalar field remains within the safe regime.  But we would like to highlight the danger that some theories of massive gravity may suffer from transplanckian variations in the value of the scalar field.
As we are agnostic about the form of the potential for the scalar field we will leave the value of $\phi_0$ arbitrary from this point but impose that $|\phi_0/\Lambda|\ll1$.

We now perturb $\phi$ and $A_{\mu}$ about these background values
\begin{eqnarray}
\phi & \rightarrow & \phi_0+\phi\;,\\
A_{\mu} & \rightarrow & \frac{1}{2}\delta_{\mu i}\epsilon_{ijk}B_jx_k+ A_{\mu}\;.
\end{eqnarray}
The equation of motion for scalar perturbations is
\begin{equation}
\Box\phi\left(1+\frac{B^2}{M^4}\right)-\frac{2}{M^4}(\nabla^2\phi B^2 -\partial_i\partial_j\phi B^iB^j)=m^2\phi -\frac{2}{\Lambda}\epsilon_{ijk}B_j(\partial_kA_i-\partial_i A_k)\;,
\label{eq:eom1}
\end{equation}
where $\Box =-\partial_t^2 +\nabla^2$.  The equation for photon perturbations is
\begin{equation}
\left(1+\frac{4\phi_0}{\Lambda}\right)\Box A_{\mu}+\frac{4}{\Lambda}\delta_{\mu i} B_j\epsilon_{ijk}\partial_k \phi=0\;.
\label{eq:eom2}
\end{equation}
 We note that the scalar field and photon fluctuations do not mix unless a conformal coupling is present\footnote{Although if we included higher order interactions of the propagating fluctuations this would not be the case.}.  Therefore we will only be able to place constraints on disformal couplings when a conformal coupling is also present.  This is similar to the results of fifth force experiments, however photon experiments will give complementary constraints. 

\subsection{The propagating modes}
\label{sec:eom}
We take the  magnetic field to be oriented  in the $z$ direction, $B_i=(0,0,B)$, and choose that photons and scalar fields propagate only in the $x$ direction.  Then after Fourier transforming with $\omega$ the frequency of the fields and $k$ their momentum in the $x$ direction, the equations of motion (\ref{eq:eom1}) and (\ref{eq:eom2}) become
\begin{equation}
\left(\begin{array}{cc}
(1+b^2)(\omega^2-k^2)+2k^2b^2-m^2 & \frac{4B k}{\Lambda} \\
\frac{4B k}{\Lambda} & (1-a^2)(\omega^2-k^2)
\end{array}\right)\left(\begin{array}{c}
\phi \\
A_y
\end{array}\right) =0 \;,\label{eq:mixing}
\end{equation}
where
\begin{equation}
a=2\sqrt{\frac{-\phi_0}{\Lambda }}\;,
\end{equation}
and
\begin{equation}
b=\frac{B}{M^2}\;.
\end{equation}
 The remaining components of $A_{\mu}$ do not mix with the scalar and travel unimpeded at the speed of light and so we do not include them in our analysis.  As mentioned before if the only potential for the scalar field is a light mass term, then this leads to large values of $a$, and the breakdown of our effective field theory description.  Therefore we assume that this is not the case, and that a sensible effective field theory description of the system is possible.  This would ensure that $a\ll1$.  The parameter $b$ controls the strength of the disformal term.

It will be helpful in what follows to have a sense of what range of values of  $b$ are well motivated, particularly from the point of view of common modifications of gravity.  In most such theories there are two typical energy scales: a light mass scale  $m_{\rm grav}\sim H_0$, corresponding to a length scale of order the size of the observable universe today and a coupling scale which will typically be at or close to the Planck scale, $M_P$.   Larger masses are less well motivated as we want modifications of gravity to be relevant on the largest scales in the universe today in order to explain the current acceleration of the expansion of the universe, but on smaller scales we find that general relativity is an excellent explanation for observed phenomena and so modifications of gravity should be negligible  there.
 Taking this at face value we would consider $m\sim H_0$ and $M\sim\Lambda\sim M_P$.
    For Tesla strength magnetic fields this choice of mass and coupling scales makes   $b \ll 1$.

There is a second well motivated possibility that arises in the decoupling limit of massive gravity where the helicity-zero mode of a massive graviton appears as an additional scalar degree of freedom.  As mentioned in the introduction, in these theories the scale controlling the disformal term is $M^2 =M_p m_{\rm grav}$.
The massive gravity motivation  gives $m=m_{\rm grav}$, $\Lambda= M_P$ and $M^4=M_P^2 m_{\rm grav}^2$.  If the graviton mass is $m_{\rm grav}\sim H_0$ this means  $b\gg 1$.  Such a small mass for the graviton is technically natural; it does not receive quantum corrections because the full diffeomorphisim invariance of general relativity is restored in the limit $m_{\rm grav}\rightarrow 0$, \cite{Hinterbichler:2011tt}. In the presence of sources this may be modified because a non trivial background scalar field profile acts to suppress the coupling of scalar field fluctuations to matter.  So the parameters are rescaled as $m=m_{\rm grav}/\sqrt{Z}$, $\Lambda=\sqrt{Z}M_P$, $M^4 =Z M_P^2m_{\rm grav}^2$, where $Z$ is a background dependent dimensionless factor which is much larger than unity.

Solutions to the equation of motion (\ref{eq:mixing}) exist when the determinant of the matrix vanishes.  This imposes
\begin{equation}
\left(\frac{\omega}{k}\right)^2=\frac{1}{1+b^2}\left\{1+\frac{m^2}{2k^2}\pm \sqrt{\left(\frac{m^2}{2k^2}-b^2\right)^2+\frac{16B^2}{k^2 \Lambda^2}\frac{1+b^2}{1-a^2}}\right\}\;.
\end{equation}
It can be verified that when the mass of the scalar field is small compared to the wavelength of the light, when $B/(k\Lambda) \ll 1$ and for $a<1$ the fields are relativistic and
\begin{equation}
\omega \sim k\;.
\end{equation}

The canonically normalized fields are given by the rescalings
\begin{equation}
\phi \rightarrow \frac{\phi}{\sqrt{1+b^2}}\;,
\label{eq:phirescale}
\end{equation}
\begin{equation}
A_y \rightarrow \frac{A_y}{\sqrt{1-a^2}}\;,
\label{eq:Arescale}
\end{equation}
and the equations of motions for the canonically normalized fields are
\begin{equation}
\left(\begin{array}{cc}
(\omega^2-k^2)+\frac{2k^2b^2-m^2}{1+b^2} & \frac{4 B k}{\Lambda\sqrt{1-a^2}\sqrt{1+b^2}} \\
\frac{4B k}{\Lambda \sqrt{1+b^2}\sqrt{1-a^2}} & (\omega^2-k^2)
\end{array}\right)\left(\begin{array}{c}
\phi \\
A_y
\end{array}\right) =0 \;.
\label{eq:eomcan}
\end{equation}

To find the probability of conversion between photons and scalars as a function of the distance travelled through the magnetic field  we  write $\omega + k \sim 2\omega$ and $\omega^2-k^2 \sim 2\omega (\omega -k)$ in equation (\ref{eq:eomcan}) and Fourier transform back to real space in the $x$ direction.  Now the equation of motion is
\begin{equation}
\left[\omega -i\partial_x +\omega\left(\begin{array}{cc}
\frac{2\omega^2b^2-m^2}{2\omega(1+b^2)} & \frac{am}{\sqrt{2}\sqrt{1-a^2}\sqrt{1+b^2}} \\
\frac{am}{\sqrt{2}\sqrt{1-a^2}\sqrt{1+b^2}} & 0
 \end{array}\right)\right] \left(\begin{array}{c}
 \phi \\
 A_y
 \end{array}
 \right) =0\;.
 \end{equation}

 This system can be diagonalized and solved to give
 \begin{equation}
 \left(\begin{array}{c}
 \phi(x)\\
 A_y(x)
 \end{array}\right)= P\left(\begin{array}{cc}
 e^{-i\omega(1 +\lambda_+)x} & 0 \\
 0 &  e^{-i\omega(1 +\lambda_-)x}
 \end{array} \right) P^{-1}\left(\begin{array}{c}
 \phi(0)\\
 A_y(0)
 \end{array}
 \right)\;,
 \label{eq:mixmat}
 \end{equation}
where
\begin{equation}
P=\left(\begin{array}{cc}
\sin\vartheta & -\cos\vartheta\\
\cos\vartheta & \sin\vartheta
\end{array}\right)\;,
\end{equation}
with
\begin{equation}
\tan 2\vartheta =\frac{4 B}{\Lambda\omega} \sqrt{\frac{1+b^2}{1-a^2}}\left(\frac{m^2}{2\omega^2}-b^2\right)^{-1}\;,
\end{equation}
\begin{equation}
\lambda_{\pm}=-\lambda(\cos2\vartheta \mp 1)\;,
\label{eq:eigenvalues}
\end{equation}
and
\begin{equation}
\lambda =\frac{1}{2(1+b^2)}\left|\frac{m^2}{2\omega^2} -b^2\right|(1+\tan^2 2\vartheta)^{1/2}\;.
\end{equation}

From equation (\ref{eq:mixmat}) we can read off the probability of transition from a photon to a scalar, and from a scalar to a photon by evaluating the modulus squared of the off diagonal terms in the mixing matrix.
The  transition probability is
\begin{equation}
P_{\gamma\rightarrow\phi}=\sin^22\vartheta\sin^2\lambda\omega x\;.
\end{equation}
This reduces to the familiar expression for  axion-like particles when $b=0$.

\subsection{The strong coupling scale}

We have taken the Lagrangian in Equation (\ref{eq:lagphotscal}) as our starting point.  It is clear from this expression that additional Lagrangian operators could be generated by quantum loop corrections, however these operators would have a higher mass dimension than the ones written explicitly in Equation (\ref{eq:lagphotscal}) and so would be suppressed by additional powers of $M$ and $\Lambda$.  The strong coupling scale is the scale below which the low energy Lagrangian (\ref{eq:lagphotscal}) can be trusted as an accurate description of the system.  Above the strong coupling scale the higher dimensional operators that we have neglected become important.  It is therefore important to check that the experiments we study take place at energies below the strong coupling scale of this theory.

An estimate of the strong coupling scale in vacuum is that it should be of the same order of magnitude as the smaller of the two energy scales in our system $M$ and $\Lambda$.  As we will see, in what follows we will consider values of $M$ down to $M \sim 10^{-1} \mbox{ eV}$.  This is problematic if we want to study the mixing of the scalar with optical photons, with electron-Volt energies, as this is already above the vacuum strong coupling scale.

However in the presence of a non-trivial background the strong coupling scale changes, this is similar in spirit to what happens to the strong coupling scale of Galileon theories as described in \cite{Nicolis:2004qq}.  We have seen that when we study scalar and photon fluctuations around a non-trivial background the kinetic terms for the fluctuations acquire background dependent prefactors.  In order to recover canonical fields we have to do the rescalings in Equations (\ref{eq:phirescale}) and (\ref{eq:Arescale}).  After performing these rescalings everywhere in the Lagrangian the interaction terms between scalars and photons also acquire factors of $1/\sqrt{1-a^2}$ and  $1/\sqrt{1+b^2}$.  Therefore in the presence of a constant background magnetic field the energy scales controlling the interactions are no-longer $M$ and $\Lambda$, but instead $(1-a^2)^{1/4}(1+b^2)^{1/4} M$ and $\sqrt{1-a^2}\sqrt{1+b^2}\Lambda$.  As discussed previously $a\ll 1$ but when $b\gg 1$  the strong coupling scale $M$ is raised to  $Mb^{1/2}$.  It is straightforward to check that for Tesla strength background magnetic fields the strong coupling scale is always greater than the $\mbox{ eV}$ energies of the optical photons used in the experiments we study.  

The optical experiments we study lie well below the strong coupling scale of the theory.  However it is possible that a low strong coupling scale in the scalar sector of the theory may percolate through loop corrections into the matter sector, where a low strong coupling scale could be problematic for precision tests.  We intend to consider this question further in future work.


\section{Laboratory constraints}
\label{sec:labconstraints}
No signs of a scalar field mixing with photons have been seen to date in the laboratory.  This means that at the optical  frequencies used in these experiments the probability of conversion between the two species of particle will be small.  This requires
 \begin{equation}
\vartheta\approx \frac{2B}{\Lambda\omega} \sqrt{1+b^2}\left(\frac{m^2}{2\omega^2}-b^2\right)^{-1}\ll 1\;,
\end{equation}
where we have simplified the expression slightly by assuming $a\ll1 $.  We know that when $b=0$ we recover the standard axion-like particle case. Indeed, the field behaves like an axion-like particle whenever $b\ll m/\omega \ll1 $ with
\begin{equation}
\vartheta \approx \frac{4B\omega}{\Lambda m^2}\;, \;\;\; \lambda \approx \frac{m^2}{4\omega^2}\;.
\end{equation}

\begin{figure}[!htb]
\centering
\subfigure[Mixing Angle, $\vartheta$]{
\includegraphics[scale=0.37]{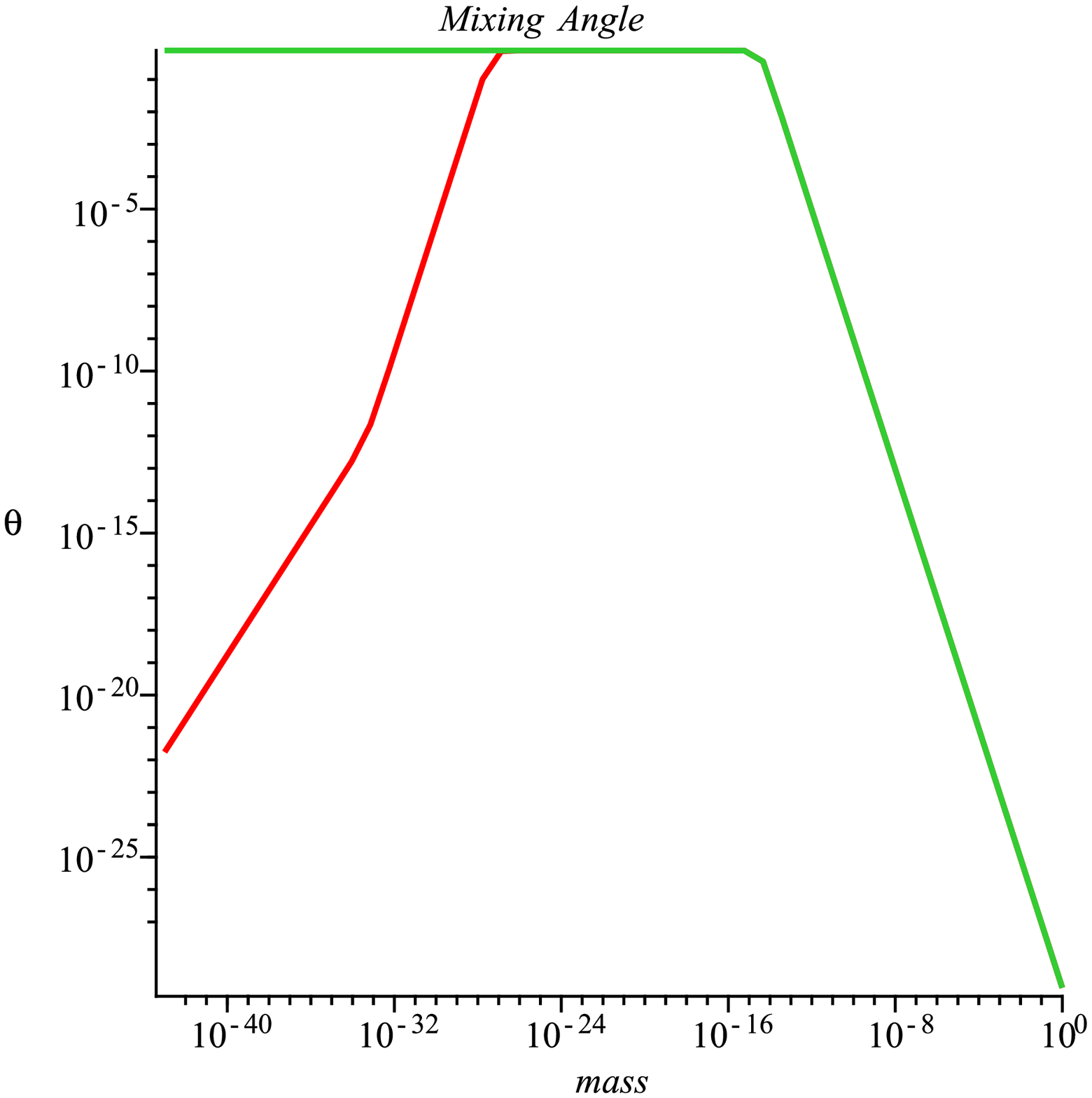}
}
\subfigure[Oscillation Wavelength, $\lambda$]{
\includegraphics[scale=0.37]{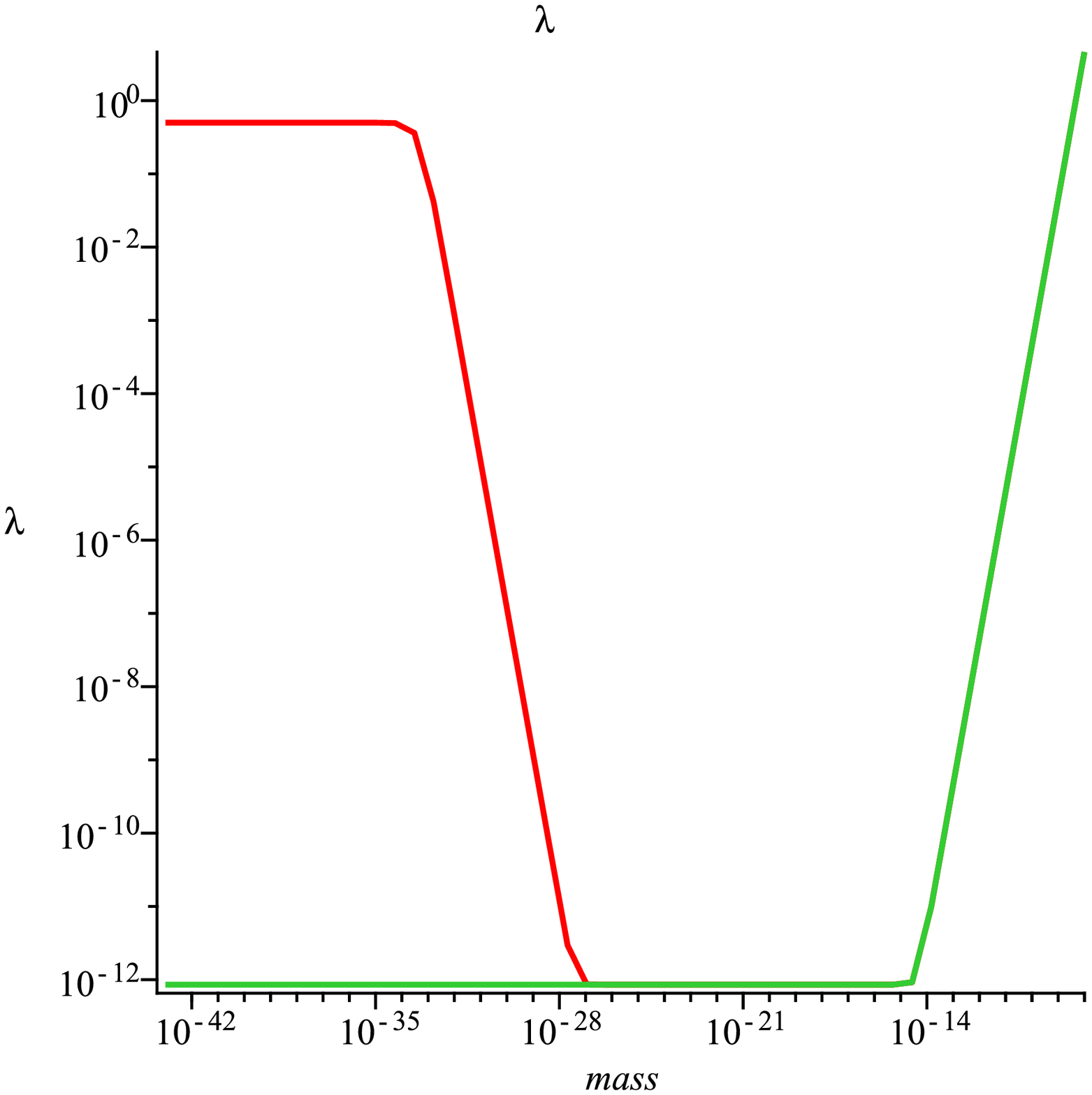}
}
\caption{The mixing parameters $\vartheta$ and $\lambda$ plotted as a function of mass $m$.  We have taken $\Lambda=10^6 \mbox{ GeV}$, $\phi_0=10^{-2}\Lambda$ and $M^2=mM_P$.  We take $B =5 \mbox{ Tesla}$ and $\omega = 2.33 \mbox{ eV}$, the experimental parameters for the ALPS experiment.  The green line shows the standard result for axion-like particles with $b=0$, the red line shows how the  effects of including a disformal coupling dominate at very low masses, which correspond to large $b$.}
\label{fig:lambdaandtheta}
\end{figure}

There are three qualitatively different regimes of behaviour that we summarise in Table \ref{tab:parameters}.
\begin{table}[tbp]
\centering

\heavyrulewidth=.08em
	\lightrulewidth=.05em
	\cmidrulewidth=.03em
	\belowrulesep=.65ex
	\belowbottomsep=0pt
	\aboverulesep=.4ex
	\abovetopsep=0pt
	\cmidrulesep=\doublerulesep
	\cmidrulekern=.5em
	\defaultaddspace=.5em
	\renewcommand{\arraystretch}{1.6}

\begin{tabular}{c c c}
$b$ & $\vartheta$ & $\lambda$ \\
\noalign{\smallskip}
\hline
\noalign{\smallskip}
$b \ll \dfrac{m}{\omega} \ll 1$ & $\dfrac{4 B \omega}{\Lambda m^2}$ & $\dfrac{m^2}{4 \omega ^2}$ \\
\noalign{\smallskip}
\hline
\noalign{\smallskip}
$\dfrac{m}{\omega}\ll b \ll 1$ & $-\dfrac{2 B}{b^2 \Lambda \omega}$ & $\dfrac{b^2}{2}$ \\
\noalign{\smallskip}
\hline
\noalign{\smallskip}
$\dfrac{m}{\omega} \ll 1 \ll b$ & $-\dfrac{2 B}{b \Lambda \omega}$ & $\dfrac{1}{2}$
\end{tabular}
\caption{Values of the mixing parameters $\vartheta$ and $\lambda$ for a range of values of $b$, the parameter that controls the relevance of the disformal terms.}
\label{tab:parameters}
\end{table}
We see that when the effects of the disformal mixing term become relevant, they act to suppress $\vartheta$ but to increase $\lambda$. For a fixed $\Lambda$ and taking the modified gravity choice $M^2=mM_P$ the variation of the mixing parameters are plotted as a function of the scalar mass in Figure \ref{fig:lambdaandtheta}, and these three regimes of behaviour are clearly visible. The suppression of $\vartheta$ means that for a given magnetic field strength and photon frequency, disformal mixing will be harder to see than standard axion-like particle behaviour.  The increase in $\lambda$ means that the length scales over which the mixing between photons and scalars oscillates is much shorter for disformal couplings.  This means that if  mixing between photons and scalars is detected, it will be straightforward to determine whether disformal effects are dominant or not.

\subsection{Light shining through a wall}
\label{sec:ALPS}

One of the most unusual consequences of scalar fields that interact with photons is their ability to make light appear to travel through opaque objects.  This is possible if a magnetic field is present on either side of the wall, allowing incoming photons to convert into scalar fields. Unlike photons these scalars can pass through the wall and then in the magnetic field on the far side of the wall they reconvert from scalars back into photons which can then be detected.

The absence of a detection of light shining through a wall has previously been used to constrain axion-like particles, but clearly it also constrains the disformal scalar fields that we consider here.  The best current constraints on light shining through wall effects come from the ALPS experiment \cite{Ehret:2010mh}.  A laser with frequency $\omega =2.33 \mbox{ eV}$ is shone through a $5\mbox{ Tesla}$ magnetic field which extends for $4.3\mbox{ m}$ before the wall and the same distance after the wall.  The experiments constrain the probability of conversion of individual  photons over $4.3\mbox{ m}$ to be
\begin{equation}
P < 2.08\times 10^{-25}\;,
\end{equation}
at $95\%$ confidence.

\begin{figure}[!htb]
\centering
\includegraphics[scale=.37]{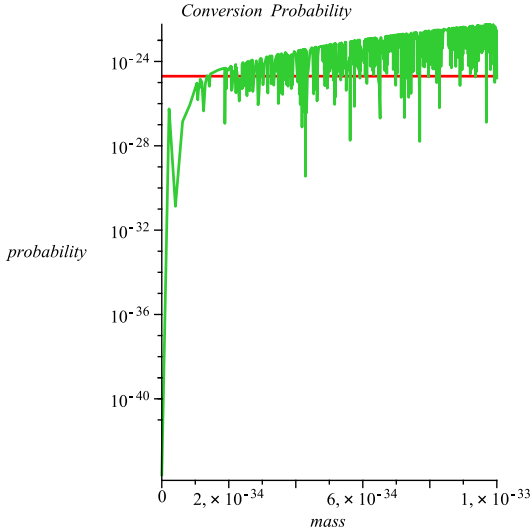}
\caption{The probability of mixing as a function of mass, for small masses where the effects of the disformal term are relevant.  We have taken $\Lambda=10^6 \mbox{ GeV}$, $\phi_0=10^{-2}\Lambda$ and $M^2=mM_P$.  The experimental parameters we take are  $B =5 \mbox{ Tesla}$, $\omega = 2.33 \mbox{ eV}$ and the distance travelled through the magnetic field is $4.3\mbox{ m}$. The red line shows the constraint of the ALPS experiment.  }
\label{fig:digraph}
\end{figure}

In Figure \ref{fig:digraph} we plot the probability of mixing as a function of the mass of the field for a typical choice of parameters.  It is clear that the effect of the disformal term is to lower the probability of mixing between scalars and photons at very low masses, and so these experiments are less constraining for disformally coupled scalars than they are for standard axion-like particles.

\begin{figure}[!htb]
\centering
\includegraphics[scale=.8]{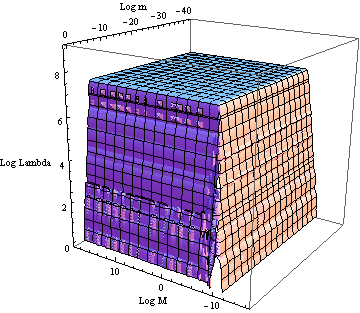}
\caption{The constraint of the ALPS experiment on the $m$, $M$, $\Lambda$ parameter space.  All regions below the surface  are excluded. The parameters are measured in  units of GeV. }
\label{fig:3dconstraint}
\end{figure}

   In Figure \ref{fig:3dconstraint} the constraint of the ALPS experiment is shown in the three dimensional parameter space $(m,M,\Lambda)$.  We see that in almost all of the interesting range the constraint on $\Lambda$ is that of the conformally coupled axion-like particle case $\Lambda \gtrsim 10^7\mbox{ GeV}$.  Constraints on $m$ and $M$ are almost as independent of the other parameters, we see that for smaller values of $\Lambda$, corresponding to larger conformal couplings, we  require $M \lesssim 10^{-11}\mbox{ GeV}$.  The constraint on $M$   reaches the most motivated choice for massive gravity $M=\sqrt{M_pH_0} = 3\times 10^{-11}\mbox{ GeV}$.  However for weaker conformal couplings  $\Lambda\gtrsim 10^7 \mbox{ GeV}$ the experiment is not able to constrain the disformal coupling $M$.

\subsection{Rotation and ellipticity}
\label{sec:PVLAS}

Assuming that the  mixing between photons and the scalar fields is small we can ask what are the effects of this mixing  on the propagation of photons through a magnetic field.  Assuming that we start with a beam composed only of photons, after traveling a distance $x$ through the magnetic field the photon wave is described by
\begin{equation}
A_{\gamma}=\cos^2\vartheta e^{-i\omega(1+\lambda_{+})x}+\sin^2\vartheta e^{-i\omega(1+\lambda_{-})x}\;.
\label{eq:pol}
\end{equation}
 Then when the photons are observed at distance $x$   the real part of the expression in equation (\ref{eq:pol}) is measured.  Writing this as $A\cos(\omega x+\delta x)$, the parameters $A$ and $\delta$ describe the change in amplitude and phase respectively.
The change is phase is
\begin{equation}
\delta x =-\lambda\omega x \cos 2\vartheta+\arctan \left[\cos2\vartheta\tan\lambda\omega x\right]\;,
\end{equation}
when the probability of mixing is small $\vartheta\ll 1$, this is approximately
\begin{equation}
\delta x \approx 2\vartheta^2 (\lambda\omega x-\tan \lambda \omega x)\;.
\end{equation}
The amplitude is
\begin{equation}
A=\cos\lambda\omega x\sqrt{1+\cos^22\vartheta\tan^2\lambda\omega x}\;,
\end{equation}
when the probability of mixing is small $\vartheta \ll 1$ this is approximately
\begin{equation}
A\approx 1-\vartheta^2\sin^2\lambda\omega x\;.
\end{equation}
Dividing the attenuation $1-A$ and the phase shift $\delta x$ by a factor of $2$ we get the rotation and the ellipticity of the  laser beam.  These are the variables measured experimentally.

The best constraints on the polarisation of light propagating through a magnetic field come from the PVLAS experiment \cite{Zavattini:2007ee}.  In this experiment light at $\omega = 1.17 \mbox{ eV}$ is passed through a $1\mbox{ m}$ length magnetic field of strength $2.3 \mbox{ Tesla}$.
No presence of a rotation signal is observed  down to
\begin{equation}
\frac{|1-A|}{2} < 1.0 \times 10^{-8}\mbox{ rad}\;,
\end{equation}
at $95\%$ confidence and no ellipticity down to
\begin{equation}
\Psi=\frac{\delta x}{2} < 1.4 \times 10^{-8} \;,
\end{equation}
at $95\%$ confidence.

\begin{figure}[!htb]
\centering
\subfigure[Rotation]{
\includegraphics[scale=0.37]{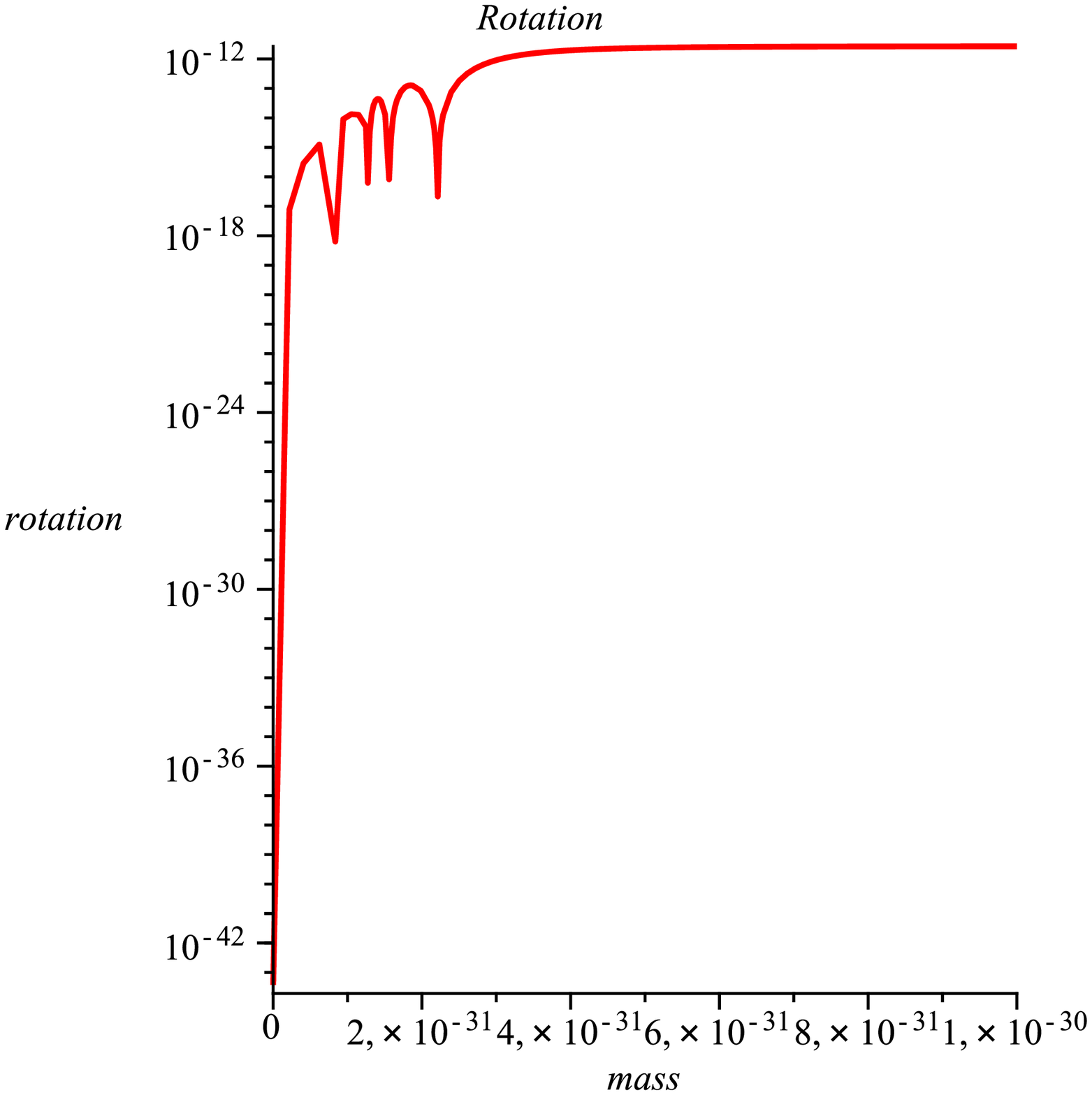}
}
\subfigure[Ellipticity]{
\includegraphics[scale=0.37]{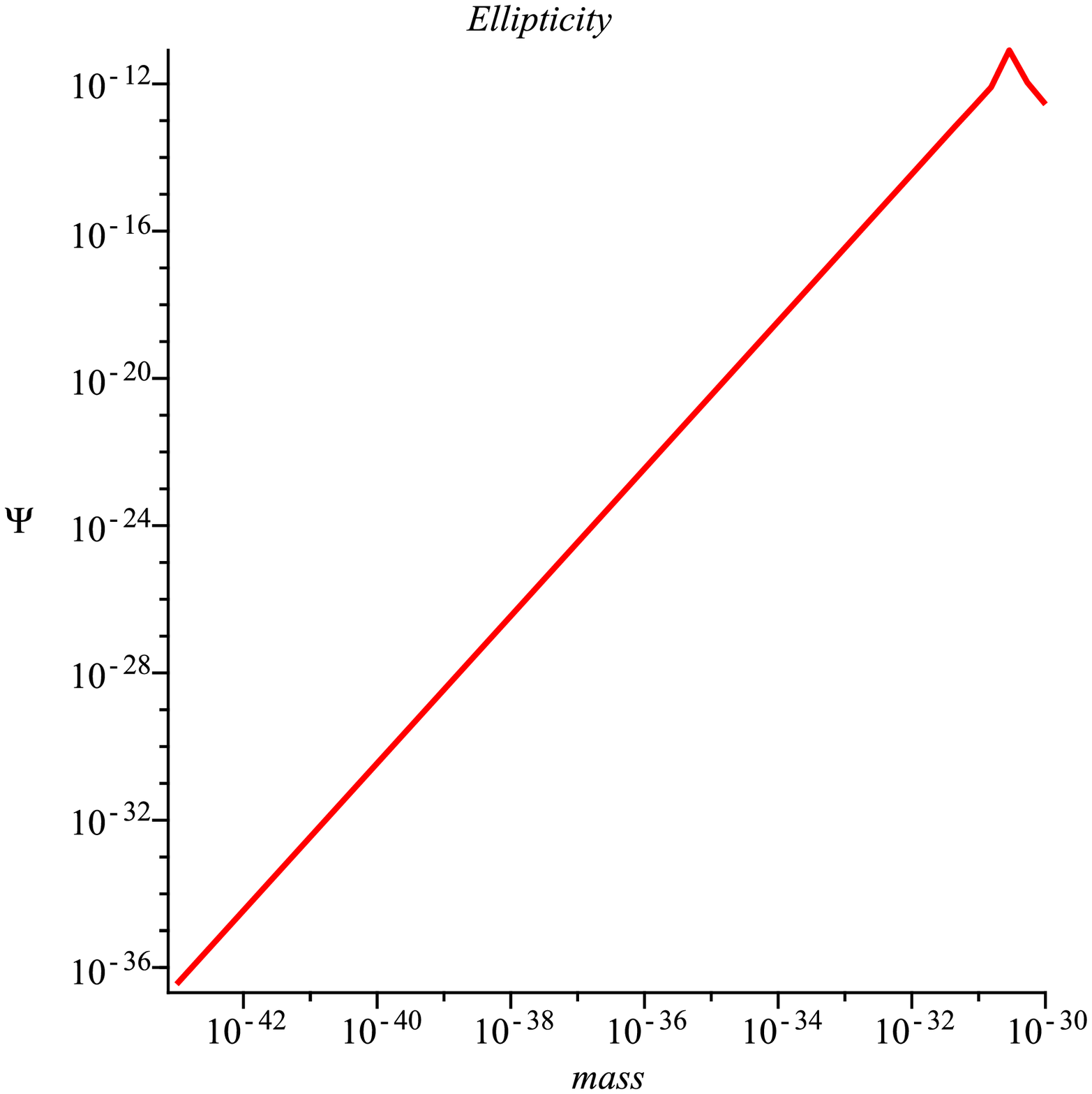}
}
\caption{The induced rotation and ellipticity $\Psi$ as a function of mass, for the PVLAS experiment.  We have taken $\Lambda=10^6 \mbox{ GeV}$, $\phi_0=10^{-2}\Lambda$ and $M^2=mM_P$.  The experimental parameters we take are  $B =2.3 \mbox{ Tesla}$, $\omega = 1.17 \mbox{ eV}$ and the distance travelled through the magnetic field is $1\mbox{ m}$.}
\label{fig:ellipticity}
\end{figure}

Figure \ref{fig:ellipticity} shows the  ellipticity and rotation induced by mixing with a scalar field  at small masses where the effects of the disformal terms are important (for typical values of the remaining parameters).  Again it is clear that at small masses  the disformal term acts to suppress the effects of the mixing between scalars and photons.

\begin{figure}[!htb]
\centering
\subfigure[Rotation]{
\includegraphics[scale=0.8]{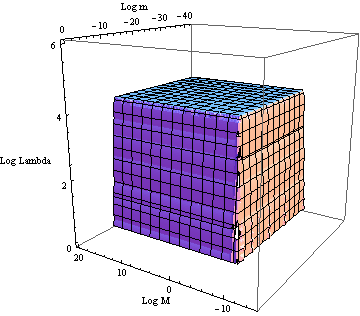}
}
\subfigure[Ellipticity]{
\includegraphics[scale=0.8]{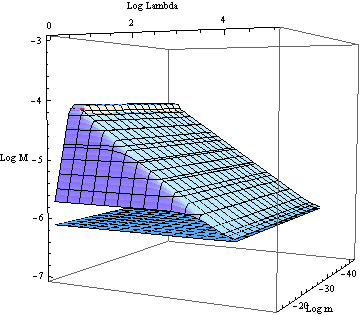}
}
\caption{The constraint of the PVLAS rotation and ellipticity measurements on the $m$, $M$, $\Lambda$ parameter space.  All regions inside the surfaces  are excluded.   All quantities are measured in units of GeV. }
\label{fig:contour_PVLAS}
\end{figure}

Figure \ref{fig:contour_PVLAS} shows the constraints of the PVLAS experiments in the $(m,M,\Lambda)$ parameter space.  We see that the measurement of ellipticity is less constraining than the measurement of rotation.  The shape of the constraint coming from the measurement of the rotation of the polarisation is similar in shape to that coming from the ALPS  experiment shown in Figure \ref{fig:3dconstraint}. We see that PVLAS is less constraining for disformally coupled scalar fields than ALPS.

\begin{figure}[!htb]
\centering
\subfigure[Rotation]{
\includegraphics[scale=0.37]{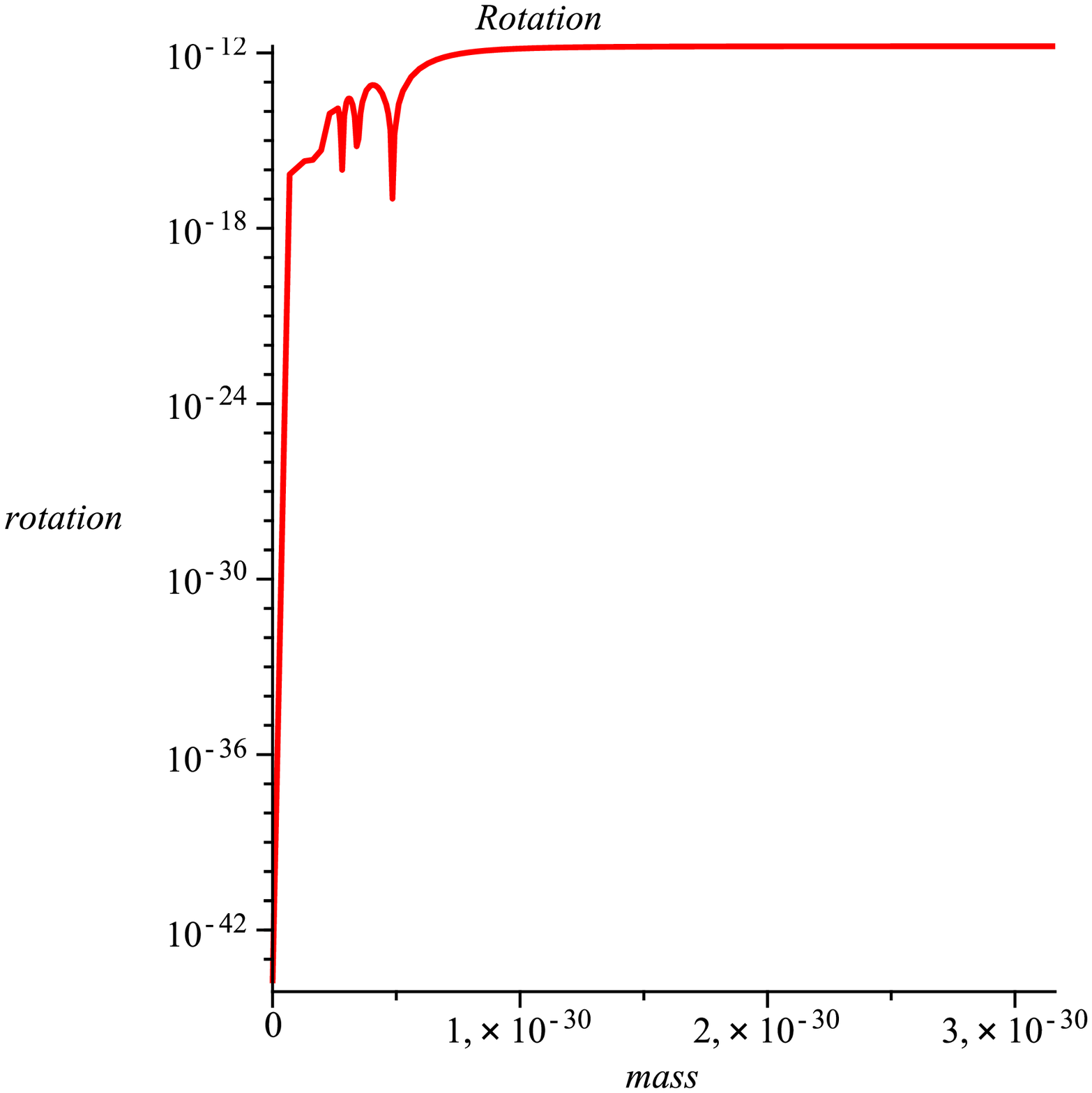}
}
\subfigure[Ellipticity]{
\includegraphics[scale=0.37]{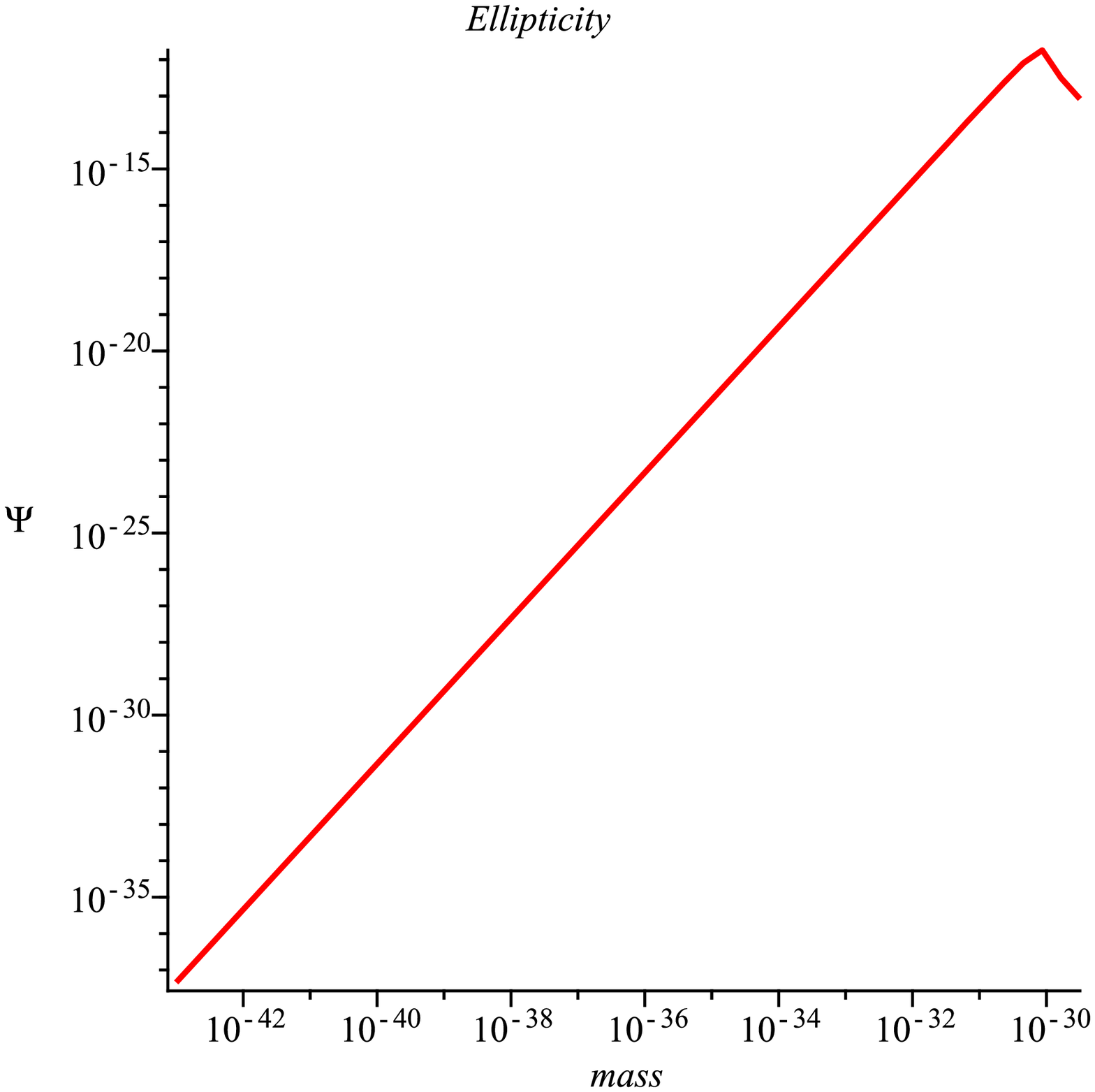}
}
\caption{The induced ellipticity $\Psi$ as a function of mass for the BMV experiment.  We have taken $\Lambda=10^6 \mbox{ GeV}$, $\phi_0=10^{-2}\Lambda$ and $M^2=mM_P$.  The experimental parameters we take are  $B =9 \mbox{ Tesla}$, $\omega = 1.17 \mbox{ eV}$ and the distance travelled through the magnetic field is $13\mbox{ cm}$. }
\label{fig:BMV}
\end{figure}

In Figure \ref{fig:BMV} we also plot the ellipticity and rotation for the parameters of the BMV experiment at the Laboratoire National des Champs Magn\'{e}tiques Intenses (LNCMI) \cite{PhysRevA.85.013837}, which is expected to release new results soon.  The BMV experiment passes a laser beam with energy $1.17\mbox{ eV}$ through $13 \mbox{ cm}$ of a $9\mbox{ T}$ magnetic field on average (its maximal value can reach $14 \mbox{T}$).  This will allow a comparison with the new measurements when they become available.

\section{Conclusions}
\label{sec:conc}

Disformal couplings between scalar fields and matter have many motivations particularly in the field of modified gravity.  However unlike most other modifications of gravity it is difficult to search for disformal couplings with static non-relativistic fifth force experiments.  In this paper we have shown that high-precision low-energy photon experiments offer a complimentary opportunity to search for and constrain disformal couplings in the laboratory.

In the region of parameter space interesting for theories of modified gravity the ALPS experiment, which looks for the effects of scalar fields mixing with photons by trying to shine light through walls, is the most constraining.  We have shown that for strong conformal couplings the null results of this experiment constrain the parameter that controls the strength of the disformal coupling to be $M\lesssim 10^{-11}\mbox{GeV}$, which reaches  the massive gravity coupling strength $M\sim (M_P H_0)^{1/2}$.  However for weaker conformal couplings the experiment is not able to constrain the disformal coupling.  Unfortunately when the disformal terms dominate the mixing between scalar fields and photons they act to suppress the strength of the mixing, which means that significant improvement in experimental accuracy would be needed to enable us to detect these fields.

The mixing between axion-like particles and photons also leads to interesting astrophysical and cosmological  effects, as the properties of light from astrophysical sources are
altered when photons propagate through  galactic, intracluster or intergalactic magnetic fields.  It will be interesting to explore the observational consequences of disformal couplings in these environments and at these frequencies, and we intend to return to this topic in future work.

\section{Acknowledgments}
We would like to thank Tony Padilla, Claudia de Rham and Javier Redondo for very helpful discussions during the preparation of this article.  CB is supported by a University of Nottingham Anne McLaren Fellowship, ACD is supported in part 
by STFC.

	\bibliographystyle{JHEPmodplain}
	\bibliography{arxiv_resubmit}

\providecommand{\href}[2]{#2}\begingroup\raggedright\begin{thebibliography}{10}

\bibitem{Bekenstein:1992pj}
J.~D. Bekenstein, {\it {The Relation between physical and gravitational
  geometry}},  {\sl Phys.Rev.} {\bf D48} (1993) 3641--3647,
  [\href{http://arxiv.org/abs/gr-qc/9211017}{{\sf arXiv:gr-qc/9211017}}],
  [\href{http://dx.doi.org/10.1103/PhysRevD.48.3641}{{\sf
  doi:10.1103/PhysRevD.48.3641}}].

\bibitem{Nicolis:2008in}
A.~Nicolis, R.~Rattazzi, and E.~Trincherini, {\it {The Galileon as a local
  modification of gravity}},  {\sl Phys.Rev.} {\bf D79} (2009) 064036,
  [\href{http://arxiv.org/abs/0811.2197}{{\sf arXiv:0811.2197}}],
  [\href{http://dx.doi.org/10.1103/PhysRevD.79.064036}{{\sf
  doi:10.1103/PhysRevD.79.064036}}].

\bibitem{Adelberger:2009zz}
E.~Adelberger, J.~Gundlach, B.~Heckel, S.~Hoedl, and S.~Schlamminger, {\it
  {Torsion balance experiments: A low-energy frontier of particle physics}},
  {\sl Prog.Part.Nucl.Phys.} {\bf 62} (2009) 102--134,
  [\href{http://dx.doi.org/10.1016/j.ppnp.2008.08.002}{{\sf
  doi:10.1016/j.ppnp.2008.08.002}}].

\bibitem{Khoury:2003rn}
J.~Khoury and A.~Weltman, {\it {Chameleon cosmology}},  {\sl Phys.Rev.} {\bf
  D69} (2004) 044026, [\href{http://arxiv.org/abs/astro-ph/0309411}{{\sf
  arXiv:astro-ph/0309411}}],
  [\href{http://dx.doi.org/10.1103/PhysRevD.69.044026}{{\sf
  doi:10.1103/PhysRevD.69.044026}}].

\bibitem{Brax:2004qh}
P.~Brax, C.~van~de Bruck, A.-C. Davis, J.~Khoury, and A.~Weltman, {\it
  {Detecting dark energy in orbit - The Cosmological chameleon}},  {\sl
  Phys.Rev.} {\bf D70} (2004) 123518,
  [\href{http://arxiv.org/abs/astro-ph/0408415}{{\sf arXiv:astro-ph/0408415}}],
  [\href{http://dx.doi.org/10.1103/PhysRevD.70.123518}{{\sf
  doi:10.1103/PhysRevD.70.123518}}].

\bibitem{Pietroni:2005pv}
M.~Pietroni, {\it {Dark energy condensation}},  {\sl Phys.Rev.} {\bf D72}
  (2005) 043535, [\href{http://arxiv.org/abs/astro-ph/0505615}{{\sf
  arXiv:astro-ph/0505615}}],
  [\href{http://dx.doi.org/10.1103/PhysRevD.72.043535}{{\sf
  doi:10.1103/PhysRevD.72.043535}}].

\bibitem{Olive:2007aj}
K.~A. Olive and M.~Pospelov, {\it {Environmental dependence of masses and
  coupling constants}},  {\sl Phys.Rev.} {\bf D77} (2008) 043524,
  [\href{http://arxiv.org/abs/0709.3825}{{\sf arXiv:0709.3825}}],
  [\href{http://dx.doi.org/10.1103/PhysRevD.77.043524}{{\sf
  doi:10.1103/PhysRevD.77.043524}}].

\bibitem{Hinterbichler:2010es}
K.~Hinterbichler and J.~Khoury, {\it {Symmetron Fields: Screening Long-Range
  Forces Through Local Symmetry Restoration}},  {\sl Phys.Rev.Lett.} {\bf 104}
  (2010) 231301, [\href{http://arxiv.org/abs/1001.4525}{{\sf
  arXiv:1001.4525}}],
  [\href{http://dx.doi.org/10.1103/PhysRevLett.104.231301}{{\sf
  doi:10.1103/PhysRevLett.104.231301}}].

\bibitem{Damour:1994zq}
T.~Damour and A.~M. Polyakov, {\it {The String dilaton and a least coupling
  principle}},  {\sl Nucl.Phys.} {\bf B423} (1994) 532--558,
  [\href{http://arxiv.org/abs/hep-th/9401069}{{\sf arXiv:hep-th/9401069}}],
  [\href{http://dx.doi.org/10.1016/0550-3213(94)90143-0}{{\sf
  doi:10.1016/0550-3213(94)90143-0}}].

\bibitem{Brax:2010gi}
P.~Brax, C.~van~de Bruck, A.-C. Davis, and D.~Shaw, {\it {The Dilaton and
  Modified Gravity}},  {\sl Phys.Rev.} {\bf D82} (2010) 063519,
  [\href{http://arxiv.org/abs/1005.3735}{{\sf arXiv:1005.3735}}],
  [\href{http://dx.doi.org/10.1103/PhysRevD.82.063519}{{\sf
  doi:10.1103/PhysRevD.82.063519}}].

\bibitem{Deffayet:2009wt}
C.~Deffayet, G.~Esposito-Farese, and A.~Vikman, {\it {Covariant Galileon}},
  {\sl Phys.Rev.} {\bf D79} (2009) 084003,
  [\href{http://arxiv.org/abs/0901.1314}{{\sf arXiv:0901.1314}}],
  [\href{http://dx.doi.org/10.1103/PhysRevD.79.084003}{{\sf
  doi:10.1103/PhysRevD.79.084003}}].

\bibitem{Vainshtein:1972sx}
A.~Vainshtein, {\it {To the problem of nonvanishing gravitation mass}},  {\sl
  Phys.Lett.} {\bf B39} (1972) 393--394,
  [\href{http://dx.doi.org/10.1016/0370-2693(72)90147-5}{{\sf
  doi:10.1016/0370-2693(72)90147-5}}].

\bibitem{Deffayet:2001uk}
C.~Deffayet, G.~Dvali, G.~Gabadadze, and A.~I. Vainshtein, {\it
  {Nonperturbative continuity in graviton mass versus perturbative
  discontinuity}},  {\sl Phys.Rev.} {\bf D65} (2002) 044026,
  [\href{http://arxiv.org/abs/hep-th/0106001}{{\sf arXiv:hep-th/0106001}}],
  [\href{http://dx.doi.org/10.1103/PhysRevD.65.044026}{{\sf
  doi:10.1103/PhysRevD.65.044026}}].

\bibitem{deRham:2010ik}
C.~de~Rham and G.~Gabadadze, {\it {Generalization of the Fierz-Pauli Action}},
  {\sl Phys.Rev.} {\bf D82} (2010) 044020,
  [\href{http://arxiv.org/abs/1007.0443}{{\sf arXiv:1007.0443}}],
  [\href{http://dx.doi.org/10.1103/PhysRevD.82.044020}{{\sf
  doi:10.1103/PhysRevD.82.044020}}].

\bibitem{deRham:2010eu}
C.~de~Rham and A.~J. Tolley, {\it {DBI and the Galileon reunited}},  {\sl JCAP}
  {\bf 1005} (2010) 015, [\href{http://arxiv.org/abs/1003.5917}{{\sf
  arXiv:1003.5917}}],
  [\href{http://dx.doi.org/10.1088/1475-7516/2010/05/015}{{\sf
  doi:10.1088/1475-7516/2010/05/015}}].

\bibitem{Brax:2012hm}
P.~Brax, {\it {Lorentz Invariance Violation in Modified Gravity}},  {\sl
  Phys.Lett.} {\bf B712} (2012) 155--160,
  [\href{http://arxiv.org/abs/1202.0740}{{\sf arXiv:1202.0740}}].

\bibitem{Noller:2012sv}
J.~Noller, {\it {Derivative Chameleons}},
  \href{http://arxiv.org/abs/1203.6639}{{\sf arXiv:1203.6639}}.

\bibitem{Kaloper:2003yf}
N.~Kaloper, {\it {Disformal inflation}},  {\sl Phys.Lett.} {\bf B583} (2004)
  1--13, [\href{http://arxiv.org/abs/hep-ph/0312002}{{\sf
  arXiv:hep-ph/0312002}}],
  [\href{http://dx.doi.org/10.1016/j.physletb.2004.01.005}{{\sf
  doi:10.1016/j.physletb.2004.01.005}}].

\bibitem{Koivisto:2008ak}
T.~S. Koivisto, {\it {Disformal quintessence}},
  \href{http://arxiv.org/abs/0811.1957}{{\sf arXiv:0811.1957}}.

\bibitem{Koivisto:2012za}
T.~S. Koivisto, D.~F. Mota, and M.~Zumalacarregui, {\it {Screening
  Modifications of Gravity through Disformally Coupled Fields}},
  \href{http://arxiv.org/abs/1205.3167}{{\sf arXiv:1205.3167}}.

\bibitem{Clayton:1999zs}
M.~Clayton and J.~Moffat, {\it {Scalar tensor gravity theory for dynamical
  light velocity}},  {\sl Phys.Lett.} {\bf B477} (2000) 269--275,
  [\href{http://arxiv.org/abs/gr-qc/9910112}{{\sf arXiv:gr-qc/9910112}}],
  [\href{http://dx.doi.org/10.1016/S0370-2693(00)00192-1}{{\sf
  doi:10.1016/S0370-2693(00)00192-1}}].

\bibitem{Drummond:1999ut}
I.~Drummond, {\it {Variable light cone theory of gravity}},
  \href{http://arxiv.org/abs/gr-qc/9908058}{{\sf arXiv:gr-qc/9908058}}.

\bibitem{Magueijo:2003gj}
J.~Magueijo, {\it {New varying speed of light theories}},  {\sl
  Rept.Prog.Phys.} {\bf 66} (2003) 2025,
  [\href{http://arxiv.org/abs/astro-ph/0305457}{{\sf arXiv:astro-ph/0305457}}],
  [\href{http://dx.doi.org/10.1088/0034-4885/66/11/R04}{{\sf
  doi:10.1088/0034-4885/66/11/R04}}].

\bibitem{Wyman:2011mp}
M.~Wyman, {\it {Galilean-invariant scalar fields can strengthen gravitational
  lensing}},  {\sl Phys.Rev.Lett.} {\bf 106} (2011) 201102,
  [\href{http://arxiv.org/abs/1101.1295}{{\sf arXiv:1101.1295}}],
  [\href{http://dx.doi.org/10.1103/PhysRevLett.106.201102}{{\sf
  doi:10.1103/PhysRevLett.106.201102}}].

\bibitem{Sjors:2011iv}
S.~Sjors and E.~Mortsell, {\it {Spherically Symmetric Solutions in Massive
  Gravity and Constraints from Galaxies}},
  \href{http://arxiv.org/abs/1111.5961}{{\sf arXiv:1111.5961}}. 20 pages, 3
  figures.

\bibitem{Babichev:2007dw}
E.~Babichev, V.~Mukhanov, and A.~Vikman, {\it {k-Essence, superluminal
  propagation, causality and emergent geometry}},  {\sl JHEP} {\bf 0802} (2008)
  101, [\href{http://arxiv.org/abs/0708.0561}{{\sf arXiv:0708.0561}}],
  [\href{http://dx.doi.org/10.1088/1126-6708/2008/02/101}{{\sf
  doi:10.1088/1126-6708/2008/02/101}}].

\bibitem{Burrage:2011cr}
C.~Burrage, C.~de~Rham, L.~Heisenberg, and A.~J. Tolley, {\it {Chronology
  Protection in Galileon Models and Massive Gravity}},  {\sl JCAP} {\bf 1207}
  (2012) 004, [\href{http://arxiv.org/abs/1111.5549}{{\sf arXiv:1111.5549}}],
  [\href{http://dx.doi.org/10.1088/1475-7516/2012/07/004}{{\sf
  doi:10.1088/1475-7516/2012/07/004}}].

\bibitem{Brax:2010uq}
P.~Brax, C.~Burrage, A.-C. Davis, D.~Seery, and A.~Weltman, {\it {Anomalous
  coupling of scalars to gauge fields}},  {\sl Phys.Lett.} {\bf B699} (2011)
  5--9, [\href{http://arxiv.org/abs/1010.4536}{{\sf arXiv:1010.4536}}],
  [\href{http://dx.doi.org/10.1016/j.physletb.2011.03.047}{{\sf
  doi:10.1016/j.physletb.2011.03.047}}].

\bibitem{Raffelt:1987im}
G.~Raffelt and L.~Stodolsky, {\it {Mixing of the Photon with Low Mass
  Particles}},  {\sl Phys.Rev.} {\bf D37} (1988) 1237,
  [\href{http://dx.doi.org/10.1103/PhysRevD.37.1237}{{\sf
  doi:10.1103/PhysRevD.37.1237}}].

\bibitem{Redondo:2010dp}
J.~Redondo and A.~Ringwald, {\it {Light shining through walls}},  {\sl
  Contemp.Phys.} {\bf 52} (2011) 211--236,
  [\href{http://arxiv.org/abs/1011.3741}{{\sf arXiv:1011.3741}}],
  [\href{http://dx.doi.org/10.1080/00107514.2011.563516}{{\sf
  doi:10.1080/00107514.2011.563516}}]. 30 pages, 16 figures.

\bibitem{Zavattini:2007ee}
{\bf PVLAS Collaboration} Collaboration, E.~Zavattini {\em et~al.}, {\it {New
  PVLAS results and limits on magnetically induced optical rotation and
  ellipticity in vacuum}},  {\sl Phys.Rev.} {\bf D77} (2008) 032006,
  [\href{http://arxiv.org/abs/0706.3419}{{\sf arXiv:0706.3419}}],
  [\href{http://dx.doi.org/10.1103/PhysRevD.77.032006}{{\sf
  doi:10.1103/PhysRevD.77.032006}}].

\bibitem{Ehret:2010mh}
K.~Ehret, M.~Frede, S.~Ghazaryan, M.~Hildebrandt, E.-A. Knabbe, {\em et~al.},
  {\it {New ALPS Results on Hidden-Sector Lightweights}},  {\sl Phys.Lett.}
  {\bf B689} (2010) 149--155, [\href{http://arxiv.org/abs/1004.1313}{{\sf
  arXiv:1004.1313}}],
  [\href{http://dx.doi.org/10.1016/j.physletb.2010.04.066}{{\sf
  doi:10.1016/j.physletb.2010.04.066}}].

\bibitem{Jaeckel:2010ni}
J.~Jaeckel and A.~Ringwald, {\it {The Low-Energy Frontier of Particle
  Physics}},  {\sl Ann.Rev.Nucl.Part.Sci.} {\bf 60} (2010) 405--437,
  [\href{http://arxiv.org/abs/1002.0329}{{\sf arXiv:1002.0329}}],
  [\href{http://dx.doi.org/10.1146/annurev.nucl.012809.104433}{{\sf
  doi:10.1146/annurev.nucl.012809.104433}}].

\bibitem{Hinterbichler:2011tt}
K.~Hinterbichler, {\it {Theoretical Aspects of Massive Gravity}},  {\sl
  Rev.Mod.Phys.} {\bf 84} (2012) 671--710,
  [\href{http://arxiv.org/abs/1105.3735}{{\sf arXiv:1105.3735}}].

\bibitem{Nicolis:2004qq}
A.~Nicolis and R.~Rattazzi, {\it {Classical and quantum consistency of the DGP
  model}},  {\sl JHEP} {\bf 0406} (2004) 059,
  [\href{http://arxiv.org/abs/hep-th/0404159}{{\sf arXiv:hep-th/0404159}}],
  [\href{http://dx.doi.org/10.1088/1126-6708/2004/06/059}{{\sf
  doi:10.1088/1126-6708/2004/06/059}}].

\bibitem{PhysRevA.85.013837}
P.~Berceau, M.~Fouch\'e, R.~Battesti, and C.~Rizzo, {\it Magnetic linear
  birefringence measurements using pulsed fields},  {\sl Phys. Rev. A} {\bf 85}
  (Jan, 2012) 013837, [\href{http://dx.doi.org/10.1103/PhysRevA.85.013837}{{\sf
  doi:10.1103/PhysRevA.85.013837}}].

\end{thebibliography}\endgroup

\end{document}